\documentstyle[12pt]{article}
\def\journal{\topmargin .3in	\oddsidemargin .5in
	\headheight 0pt	\headsep 0pt
	\textwidth 5.625in % 1.2 preprint size  %6.5in	
	\textheight 8.25in % 1.2 preprint size 9in
	\marginparwidth 1.5in
	\parindent 2em
	\parskip .5ex plus .1ex		\jot = 1.5ex}
\journal

\catcode`\@=11
\def\marginnote#1{}
\newcount\hour
\newcount\minute
\newtoks\amorpm
\hour=\time\divide\hour by60
\minute=\time{\multiply\hour by60 \global\advance\minute by-\hour}
\edef\standardtime{{\ifnum\hour<12 \global\amorpm={am}%
	\else\global\amorpm={pm}\advance\hour by-12 \fi
	\ifnum\hour=0 \hour=12 \fi
	\number\hour:\ifnum\minute<10 0\fi\number\minute\the\amorpm}}
\edef\militarytime{\number\hour:\ifnum\minute<10 0\fi\number\minute}
\def\draftlabel#1{{\@bsphack\if@filesw {\let\thepage\relax
   \xdef\@gtempa{\write\@auxout{\string
      \newlabel{#1}{{\@currentlabel}{\thepage}}}}}\@gtempa
   \if@nobreak \ifvmode\nobreak\fi\fi\fi\@esphack}
	\gdef\@eqnlabel{#1}}
\def\@eqnlabel{}
\def\@vacuum{}
\def\draftmarginnote#1{\marginpar{\raggedright\scriptsize\tt#1}}
\def\draft{\oddsidemargin -.5truein
	\def\@oddfoot{\sl preliminary draft \hfil 
	\rm\thepage\hfil\sl\today\quad\militarytime}
	\let\@evenfoot\@oddfoot	\overfullrule 3pt
	\let\label=\draftlabel 
	\let\marginnote=\draftmarginnote
   \def\@eqnnum{(\theequation)\rlap{\kern\marginparsep\tt\@eqnlabel}%
\global\let\@eqnlabel\@vacuum}  }
\def\preprint{\twocolumn\sloppy\flushbottom\parindent 2em
	\leftmargini 2em\leftmarginv .5em\leftmarginvi .5em
	\oddsidemargin -.5in	\evensidemargin -.5in
	\columnsep .4in	\footheight 0pt
	\textwidth 10in	\topmargin  -.4in
	\headheight 12pt \topskip .4in
	\textheight 7.1in \footskip 0pt 
	\def\@oddhead{\thepage\hfil\addtocounter{page}{1}\thepage}
	\let\@evenhead\@oddhead	\def\@oddfoot{}	\def\@evenfoot{} }
\def\numberbysection{\@addtoreset{equation}{section}
	\def\theequation{\thesection.\arabic{equation}}}
\def\underline#1{\relax\ifmmode\@@underline#1\else
	$\@@underline{\hbox{#1}}$\relax\fi}
\def\titlepage{\@restonecolfalse\if@twocolumn\@restonecoltrue\onecolumn
     \else \newpage \fi \thispagestyle{empty}\c@page\z@	
	\def\thefootnote{\fnsymbol{footnote}} }
\def\endtitlepage{\if@restonecol\twocolumn \else \newpage \fi
	\def\thefootnote{\arabic{footnote}} 
	\setcounter{footnote}{0}}  %\c@footnote\z@ }
\catcode`@=12
\relax
\def\figcap{\section*{Figure Captions\markboth
	{FIGURECAPTIONS}{FIGURECAPTIONS}}\list
	{Figure \arabic{enumi}:\hfill}{\settowidth\labelwidth{Figure 999:}
	\leftmargin\labelwidth 
	\advance\leftmargin\labelsep\usecounter{enumi}}}
 \relax
\def\tablecap{\section*{Table Captions\markboth
	{TABLECAPTIONS}{TABLECAPTIONS}}\list
	{Table \arabic{enumi}:\hfill}{\settowidth\labelwidth{Table 999:}
	\leftmargin\labelwidth 
	\advance\leftmargin\labelsep\usecounter{enumi}}}
 \relax
\def\reflist{\section*{References\markboth
	{REFLIST}{REFLIST}}\list
	{[\arabic{enumi}]\hfill}{\settowidth\labelwidth{[999]}
	\leftmargin\labelwidth 
	\advance\leftmargin\labelsep\usecounter{enumi}}}
 \relax
\makeatletter
\newcounter{pubctr}
\def\publist{\@ifnextchar[{\@publist}{\@@publist}}
\def\@publist[#1]{\list
	{[\arabic{pubctr}]\hfill}{\settowidth\labelwidth{[999]}
	\leftmargin\labelwidth 
	\advance\leftmargin\labelsep
	\@nmbrlisttrue\def\@listctr{pubctr}
	\setcounter{pubctr}{#1}\addtocounter{pubctr}{-1}}}
\def\@@publist{\list
	{[\arabic{pubctr}]\hfill}{\settowidth\labelwidth{[999]}
	\leftmargin\labelwidth 
	\advance\leftmargin\labelsep
	\@nmbrlisttrue\def\@listctr{pubctr}}}
 \relax
\makeatother
\catcode`\@=11
\def\section{\@startsection {section}{1}{0pt}{-3.5ex plus -1ex minus 
 -.2ex}{2.3ex plus .2ex}{\raggedright\large\bf}}
\catcode`\@=12
\newskip\humongous \humongous=0pt plus 1000pt minus 1000pt

\newif\ifdtup

\def\oldreffmt#1{\rlap{[#1]} \hbox to 2\parindent{}}

\def\figfmt#1{\rlap{Figure {#1}} \hbox to 1in{}}

\let\vev\VEV

\def\ltap{\raisebox{-.4ex}{\rlap{$\sim$}} \raisebox{.4ex}{$<$}}
\def\gtap{\raisebox{-.4ex}{\rlap{$\sim$}} \raisebox{.4ex}{$>$}}
\def\beq{\begin{equation}}
\def\eeq{\end{equation}}

\def\bea{\begin{eqnarray}}                                 
\def\eea{\end{eqnarray}}
\def\np#1#2#3{           {\it Nucl. Phys. }{\bf #1}, #2 (19#3)}
\def\pl#1#2#3{           {\it Phys. Lett. }{\bf #1}, #2 (19#3)}
\def\pr#1#2#3{           {\it Phys. Rev. }{\bf #1}, #2 (19#3)}

\def\prl#1#2#3{          {\it Phys. Rev. Lett. }{\bf #1}, #2 (19#3)}

\def\zp#1#2#3{           {\it Zeit. fur Physik }{\bf #1}, #2 (19#3)}
\catcode`\@=11
\def\eqnarray{\stepcounter{equation}\let\@currentlabel=\theequation
\global\@eqnswtrue
\global\@eqcnt\z@\tabskip\@centering\let\\=\@eqncr
\gdef\@@fix{}\def\eqno##1{\gdef\@@fix{##1}}%
$$\halign to \displaywidth\bgroup\@eqnsel\hskip\@centering
  $\displaystyle\tabskip\z@{##}$&\global\@eqcnt\@ne
  \hskip 2\arraycolsep \hfil${##}$\hfil
  &\global\@eqcnt\tw@ \hskip 2\arraycolsep $\displaystyle\tabskip\z@{##}$\hfil
   \tabskip\@centering&\llap{##}\tabskip\z@\cr}

\def\@@eqncr{\let\@tempa\relax
    \ifcase\@eqcnt \def\@tempa{& & &}\or \def\@tempa{& &}
      \else \def\@tempa{&}\fi
     \@tempa \if@eqnsw\@eqnnum\stepcounter{equation}\else\@@fix\gdef\@@fix{}\fi
     \global\@eqnswtrue\global\@eqcnt\z@\cr}

\catcode`\@=12
\font\tenbifull=cmmib10 % bold math italic
\font\tenbimed=cmmib10 scaled 800
\font\tenbismall=cmmib10 scaled 666
\relax

\def\np#1#2#3{        {Nucl. Phys. }{\bf #1}, #2 (19#3)}
\def\pl#1#2#3{        {Phys. Lett. }{\bf #1}, #2 (19#3)}
\def\pr#1#2#3{        {Phys. Rev. }{\bf #1}, #2 (19#3)}

\def\prl#1#2#3{       {Phys. Rev. Lett. }{\bf #1}, #2 (19#3)}
\def\arn#1#2#3{       {Ann. Rev. Nucl. Part. Sci. }{\bf #1}, #2 (19#3)}

\def\zp#1#2#3{       {Z. Phys. }{\bf #1}, #2 (19#3)}
\begin{document}
\begin{titlepage}
\today          \hfill 
\begin{center}
\hfill    LBL-38569 \\

\vskip .25in

{\large \bf A Consistent model of Electroweak data including 
$Z\to b\overline{b}$ and
$Z\to c \overline{c}$} \footnote{This work was supported in part 
by the Director, Office of Energy 
Research, Office of High Energy and Nuclear Physics, Division of High 
Energy Physics of the U.S. Department of Energy under Contract 
DE-AC03-76SF00098 and in part by the National Science Foundation 
under grant PHY-90-21139.}

\vskip .25in

\vskip .25in
K. Agashe\footnote{email:agashe@theor3.lbl.gov.}, M. Graesser
\footnote{email:graesser@theor3.lbl.gov}
\footnote{Supported by a Natural Sciences and Engineering 
Research Council of Canada Fellowship}, 
Ian Hinchliffe\footnote{email:hinchliffe@theor3.lbl.gov},  
M. Suzuki\\

{\em Theoretical Physics Group\\
    Ernest Orlando Lawrence Berkeley National Laboratory\\
      University of California\\
    Berkeley, California 94720}
\end{center}

\vskip .25in

\begin{abstract}
    We have performed an overall fit to the electroweak data with the
generation blind $U(1)$ extension of the Standard Model.  
As input data
for fitting we have included the asymmetry parameters, 
the particle decay
widths of $Z$, neutrino scattering, and atomic parity violation.  
The QCD
coupling $\alpha_s$ has been constrained to the world average 
obtained from
all data except the $Z$ width.  On the basis of our fit we have 
constructed a
viable gauge model that not only explains $R_b$ and $R_c$ but also 
provides
a much better overall fit to the data than the Standard Model. Despite 
its phenomenological
viability, our model is unfortunately not simple from the theoretical
viewpoint.  Atomic parity violation experiments strongly disfavor 
more aesthetically appealing alternatives that can be grand unified.
\end{abstract}
\end{titlepage}
%THIS PAGE (PAGE ii) CONTAINS THE LBL DISCLAIMER
%TEXT SHOULD BEGIN ON NEXT PAGE (PAGE 1)
\renewcommand{\thepage}{\roman{page}}
\setcounter{page}{2}
\mbox{ }

\vskip 1in

\begin{center}
{\bf Disclaimer}
\end{center}

\vskip .2in

\begin{scriptsize}
\begin{quotation}
This document was prepared as an account of work sponsored by the United
States Government. While this document is believed to contain correct 
 information, neither the United States Government nor any agency
thereof, nor The Regents of the University of California, nor any of their
employees, makes any warranty, express or implied, or assumes any legal
liability or responsibility for the accuracy, completeness, or usefulness
of any information, apparatus, product, or process disclosed, or represents
that its use would not infringe privately owned rights.  Reference herein
to any specific commercial products process, or service by its trade name,
trademark, manufacturer, or otherwise, does not necessarily constitute or
imply its endorsement, recommendation, or favoring by the United States
Government or any agency thereof, or The Regents of the University of
California.  The views and opinions of authors expressed herein do not
necessarily state or reflect those of the United States Government or any
agency thereof, or The Regents of the University of California.
\end{quotation}
\end{scriptsize}

\vskip 2in

\begin{center}
\begin{small}
{\it Lawrence Berkeley Laboratory is an equal opportunity employer.}
\end{small}
\end{center}

\newpage
\renewcommand{\thepage}{\arabic{page}}
\setcounter{page}{1}

The observation at LEP\cite{lepdata} 
that the decay widths of the $Z$ to
$b \overline{b}$ and $c \overline{c}$ do not agree with the 
Standard Model 
expectations \cite{smexpect} has led to a
flurry of theoretical activity \cite{susy,techni,holdom,quark,U(1),marchr}. 
Various possible explanations have been
considered. Most of these explanations suffer from at least one defect. 
Either they do not present a complete phenomenologically viable model or
they present an overall fit that ignores some other experimental data.
In this paper we present a model that, while aesthetically distasteful, is
phenomenologically viable and has a much better overall fit to data than 
the
Standard Model. As input data we use the various asymmetries and partial 
widths as measured on the $Z$ resonance as well as other data that are 
constraining.

First, we describe the philosophy of our model. Couplings of the $Z$ to 
leptons are severely constrained by the current data, so we
modify only the couplings to quarks. We modify the couplings in a 
generation
independent fashion and demonstrate the modifications needed to
obtain a good fit to the data. This fit is quantified in terms of the total
$\chi^2$. We next show that a model involving the mixing of the $Z$ with a
second more massive boson can be constructed. The model is not 
supersymmetric
and requires the existence of new quarks to ensure anomaly cancellation.
Finally we comment on the constraints that the non-observation of such 
particles and the
new gauge boson itself place on the model and conclude.

It is important that any model that purports to explain the 
problems in the $b \overline{b}$ and $c \overline{c}$ decay widths 
of the $Z$ does not introduce 
problems with other processes. Quantities that are measured precisely at 
the
$Z$ are \cite{lepdata}, the mass of the $Z$, the forward-backward asymmetry  
for leptons
($A^{\ell}_{FB}$), for charm ($A^{c}_{FB}$) and for bottom ($A^b_{FB}$) 
quarks;
the asymmetries measured in tau decay ($A_{\tau}$ and $A_e$),
the total width of the $Z$ ($\Gamma_Z$), the hadronic production 
cross section 
($\sigma_h^0$), the ratio of the hadronic to 
leptonic width  ($R_{\ell}$), 
the fraction of the hadronic width that goes into charm quarks ($R_c$)  
and bottom quarks ($R_{b}$); as well as the left-right beam 
polarization asymmetry 
($A_{LR}$) and left-right forward-backward asymmetries for 
charm ($A_{c} (LR)$) and bottom quarks ($A_{b} (LR)$) \cite{slac}.
In addition there are other important pieces of data.
The first of these is $\alpha_s$ that we constrain to be equal to the 
world
average \cite{hinch95} obtained from all data except the $Z$ width; 
we include 
the measurements from jet counting at the $Z$ \cite{jet-count} since 
these measurements are independent of the
couplings of the quarks to the $Z$ itself.  Very important are data from
lower energy experiments, particularly the measurement of parity violation 
in
cesium ($Q_W^{Cs}$) \cite{cesium} and thallium ($Q_W^{Tl}$) \cite{thalium} 
atoms which severely constrain the vector couplings of the $Z$ to up
and down quarks. The $W$ mass ($M_W$) \cite{wmass} severely constrains 
any shifts in the gauge boson mass spectra and finally measurements from 
neutral current interaction of neutrinos \cite{neutrino}
constrain the couplings of up and down quarks to the $Z$ at lower energies.
We shall include all of these data in our fit. Models that can be favored by
the $Z$ data alone are disfavored when the rest of the data are included.

The measured values of $R_b=0.2219\pm 0.0017$
and $R_c=0.1540\pm 0.0074$ deviate by  $3.67\sigma$ and
$2.46\sigma$, respectively, from the Standard Model predictions. 
The value of $R_c$ is 10\% lower than the Standard Model value. 
It is difficult to explain these discrepancies 
by models based on radiative corrections \cite{susy}.
Since $R_{q}\propto  g_L^{q2} + g_R^{q2}$, where
$g_{L(R)}^q$ is the left-handed (right-handed) coupling of the quark
to the $Z$ boson, we need shifts in these couplings due to new physics
to resolve the $R_b$ and $R_c$ anomalies.  If the new physics affects only
the $b$ and $c$ quark couplings, such shifts are difficult to 
reconcile with the otherwise good agreement with the Standard Model for the 
following reason.  Since the QCD corrections to the partial
decay widths cancel to good accuracy in $R_b$ and $R_c$,
a shift in $R_b$ and $R_c$ changes the total hadronic decay rate into
\begin{equation}
\Gamma_{had}=\Gamma_{had}^0
      \times \biggl(1+\frac{\alpha_s(M_Z)}{\pi}+O(\alpha_s^2) \biggr)\times
    \biggl(1+\frac{\delta R_b+\delta R_c}{1-R_b^{SM}-R_c^{SM}}\biggr),
\end{equation} 
where $\delta R_b\equiv R_b^{exp}- R_b^{SM}$,  
$\delta R_c\equiv R_c^{exp}-R_c^{SM}$, and $\Gamma_{had}^0$ denotes the 
Standard Model
value of $\Gamma_{had}$
before the QCD correction. With $\alpha_s(M_Z) = 0.12$, this change
would shift $\Gamma_Z$ by $-11\sigma$ from the measured value and, in
terms of $R_l$, by $-14\sigma$. 
If instead $\alpha_s(M_Z)$ is extracted by fitting 
$\Gamma_{had}$ to its measured value, $\alpha_s(M_Z)$
would have to be $0.186 \pm 0.042$, in disagreement with the world average 
of $0.118 \pm 0.003$. 
A natural resolution of this problem involves postulating the new physics 
for
other quarks too.  In particular, if the new physics is generation blind, 
the model is free from the fine tuning problem of flavor changing neutral 
currents, which is a common difficulty in the class of models 
\cite{holdom,quark} that introduce new physics only in the 
heavy generations.  

The simplest way to accommodate these features is to
add another $U(1)$ factor to $SU(2) \times U(1)_Y$ of 
the Standard Model \cite{holdom,U(1),marchr}.
Mixing between the gauge boson $X$ of this extra $U(1)$ with the $Z$ 
boson of the Standard Model can produce the shifts in the $Zq\bar{q}$ 
couplings 
that are necessary to explain $R_b$, $R_c$ and $\alpha_s$. The
most general generation-blind $U(1)_X$ current that is consistent 
with 
$SU(2) \times U(1)_Y$ can be written as  
\begin{equation}
J_X^{\mu} = g_X(q_Q\overline{Q}\gamma^\mu Q 
              +q_U\overline{U}\gamma^\mu U
              +q_D\overline{D}\gamma^\mu D
              +q_L\overline{L}\gamma^\mu L
              +q_E\overline{E}\gamma^\mu E + \cdots) ,
\end{equation} 
where $Q$ and $L$ represent the left-handed quark and lepton doublets, 
and $U$,
$D$ and $E$ are the right-handed up-type quarks, down-type quarks and
charged leptons, respectively.  Summation over generations is understood,
and the contributions from particles other than those of the Standard 
Model have been 
suppressed. Since $U(1)_X$ charges always enter multiplied by $g_X$,
we normalize them to $q_Q =-1$ so that five parameters, $g_X$ and  
four charge ratios, specify the $U(1)_X$ current.
The $Z-X$ mixing occurs by Higgs doublets that carry $U(1)_X$ charges. 
We
assume that there is no higher dimensional Higgs multiplet of $SU(2)$.  
Then
the mass matrix may be written as
\begin{equation}
M^2= 
\left( 
 \begin{array}{ll}  
\frac{1}{4} g_Z^2v^2 &       \kappa g_Zg_Xv^2\cr
      
        \kappa g_Zg_X v^2      &   g_X^2V_X^2 
\end{array}
\right)
\end{equation} 
where $g_Z = g_2/\cos\theta_W$ with $g_2$ being the $SU(2)$ coupling and 
$v^2 = (\sqrt{2}G_F)^{-1}$ if the tree level expression for the $W$ mass 
is unchanged. The gauge eigenstates ($Z$,$X$) are related to the
mass eigenstates ($Z_M,X_M$) by
\begin{equation} 
\begin{array}{rl}
Z&=Z_M \cos \alpha - X_M \sin \alpha \cr
X&=X_M \cos \alpha + Z_M \sin \alpha
\end{array}
\end{equation} 
where
$\tan 2\alpha =-2\kappa g_Zg_Xv^2/(g_X^2V_X^2-g_Z^2v^2/4)$. The coupling 
of 
the Standard Model quark of flavor {\it i} to the $Z$ gauge boson 
is given by
$J^\mu = g_Zq_i(\overline{q}\gamma^{\mu} q)$ with $q_i \equiv (T_{3L} 
- Q\sin^2 \theta_W)_i$. The lighter mass eigenstate $Z_M$ is identified 
with 
the experimentally observed $Z$ boson.
The mixing between $Z$ and $X$ shifts the
coupling to the observed $Z$ from the Standard Model value by \\
    $\delta q_i  = (g_X/g_Z)q_{Xi} \sin\alpha + q_i(\cos\alpha - 1)$, 
where $q_{Xi}$ is the $U(1)_X$ charge of quark $i$. 
If $M_X \gg M_Z$, the mixing angle is given by 
$\sin\alpha=-\kappa g_Z v^2/(g_X V_X^2)$ and therefore
$\delta q_i = -\kappa q_{Xi}(v/V_X)^2$.
In this approximation there are two parameters $\kappa, V_X$ for mixing 
and 
four $U(1)_X$ charge ratios that are fitted to the data. When $M_X$
is comparable with $M_Z$, exact diagonalization must be 
done and the gauge coupling $g_X$ is included as an independent parameter.
  
  Since the $Z$ gauge boson is not a mass eigenstate, the tree-level 
relation 
$M_Z^2=g_Z^2v^2/4$ is no longer valid. However, the mass relation
$M_W^2 = g^2v^2/4$ is not affected. The shift in $M_Z$ can
be expressed as a shift in the $\rho$ parameter \cite{rho-param} .  
Since the
$Z$ mass is measured more accurately than the $W$ mass, we use the $W$ 
mass  
relative to the $Z$ mass as experimental information in comparing 
theoretical predictions with the data. The decrease in $M_Z$ is 
translated 
into an increase in $M_W$ and a decrease in $\sin^2 \theta _W$. 
In the large 
$M_X$ approximation,  
\begin{eqnarray}
  \frac{\delta M_W^2}{M_W^2} & = & -\frac{\delta M_Z^2}{M_Z^2} 
      \frac{\cos^2 \theta _W}{\cos^2 \theta_W - \sin^2 \theta _W}, \\
  \frac{\delta \sin^2 \theta _W}{\sin^2 \theta _W}
                       & = &\frac{\delta M_Z^2}{M_Z^2} \frac{\cos^2 
\theta_W 
                     \sin^2 \theta _W}{\cos^2 \theta_W-\sin^2 \theta _W},
\end{eqnarray} 
with $\delta M_Z^2/M_Z^2 = -(M_X \sin\alpha /M_Z )^2$.

Atomic parity violation experiments constrain the vector
couplings of the up and the down type quarks. For a
heavy atom with atomic number Z and neutron number N, these experiments
measure the charge \\
$ Q_W = -2((2Z+N)C_{1u}+(2N+Z)C_{1d})$, where 
$C_{1q}$ is defined in \cite{Ci}.
The measured \cite{cesium,thalium} and predicted 
\cite{cesiumth,thaliumth} $Q_W$ 
charges for cesium (Z=55, N=78) and thallium (Z=81, N=124) are:
\begin{equation}
\begin{array}{ll}
    Q_W(Cs)=-71.04\pm 1.81,\;\;\;\; &Q_W(Cs)^{SM}=-73.14\;(1.16\sigma)\cr
    Q_W(Tl)=-114.2\pm 3.8, \;\;\;\; &Q_W(Tl)^{SM}=-116.3\;(0.55\sigma).
\end{array}
\end{equation}  
Both experiments agree on the sign of the difference between the measured 
value 
and the Standard Model prediction. These measurements strongly constrain 
any new physics that would further decrease the $Q_W$ charge and hence 
limit
the values of the $U(1)_X$ charges the quarks can have.

    We perform a minimum $\chi^2$ analysis fitting both the shifts in 
the vector and axial couplings of $Z$ and the shifts in $M_W$ and $\sin^2 
\theta _W$ to the 18 observables discussed above.
Although the SLD measurement \cite{slac} of $A_{LR}$ 
is inconsistent with the LEP measurement, 
we find no reason to exclude either measurement from the fit. 
Electroweak radiative corrections \cite{neutrino,ew-rad} are
incorporated in the Standard Model values of these observables.  
For the $Z-X$ mixing
contribution to these observables, no radiative corrections are 
included.  Although loop corrections can generate kinetic energy 
mixing between $Z$ and $X$, such mixing 
is equivalent to the mass mixing at any fixed $q^2$. It makes a small 
difference only when extrapolation is made to different $q^2$ 
which is relevant 
when fitting to the low-energy experiments. However, 
we checked that the $Z-X$ kinetic mixing parameters vary by a negligible 
amount over this range, 
and so we may ignore the extrapolation in the $U(1)_X$ current. 
In the case of the low energy
parameters, we make only the Standard Model radiative corrections by 
running from 
$q^2=M_Z^2$ to $q^2=0$. In computing the electroweak radiative corrections,
we use $\alpha_s=0.118$, $m_t=175$ GeV and $1/\alpha(M_Z)=128.75$.  
The new physics requires a nonminimal Higgs sector. The radiative 
corrections
due to Higgs loops are numerically very small.  Therefore we 
approximate the
Higgs correction with that of the Standard Model by choosing two 
values (100 GeV and 400 GeV) for the Higgs mass.
  
In performing the fit we restrict to $V_X > 750$ GeV (see later) and 
allow the leptons to 
have arbitrary $U(1)_X$ charges. We diagonalize the $Z-X$
mass matrix exactly. The minimum $\chi^2$ 
is 16 for $M_H=100$ GeV and the preferred value of the $U(1)_X$
charges of the leptons is zero.
Setting these charges to zero, we have four parameters.
With fourteen degrees of freedom, our best $\chi^2$ is
16 for $M_H=100$ GeV and 15 for $M_H=400$ GeV. 
For comparison, the $\chi^2$ for the Standard Model is 30 for 
$M_H=100$ GeV.
The $U(1)_X$ charges are 
$q_U=2.47\pm 0.33$ and $q_D=1.17\pm 0.46$ for $M_H=100$ GeV, and 
$q_U=2.19\pm 0.69$ and $q_D=0.905\pm 0.42$ for $M_H=400$ GeV. The 
errors correspond to
$\chi^2=\chi^2_{min}+1$. See Table 1 for
the experimental and fitted values of the observables.

We now build a model based on our analysis. Since the leptons 
carry no $U(1)_X$
charge, there are three logical possibilities in constructing a 
two-doublet
Higgs model: (1) $q_U=2q_L-q_D$; (2) $q_U=q_Q$; and (3) $q_D=q_Q$. 
We note that the fitted
$U(1)_X$ charges are inconsistent with these possibilities.
If the atomic parity violation data are excluded, only case (3) is 
favored by the 
remaining data; the $\chi ^2$
is 19 (17) for $m_H=$ 100 (400) GeV and for 13 degrees of freedom
(for comparison, the $\chi ^2$ for the Standard Model without these data 
is 28 for 16 degrees of
freedom). When the atomic parity violation data are included the $\chi^2$ 
is 28 for 
15 degrees of freedom. As we see no basis for ignoring these data, 
$q_Q=q_D$ is not feasible and 
we must introduce three Higgs doublets. Anomaly cancellation is 
challenging and requires many more fermions than in the Standard Model.
\footnote{Recently Babu 
{\it et al} \cite{marchr} proposed
supersymmetric U(1) extension
models for $R_b$ and $R_c$. To accommodate supersymmetry
$q_Q = q_D$ was imposed on all models.
The good $\chi^2$ that they obtained
with $q_Q = q_D$ is mainly due to the neglect of atomic parity violation.
If atomic parity violation is taken into account, the $\chi^2$ of their
models would be much larger. One of their models is specially attractive 
since
it can be embedded into the E(6) grand unified model. With atomic parity
violation, however, the $\chi^2$ is only slightly better than the 
Standard
Model.}
  
\begin{table}
\begin{center}
\begin{tabular}{||l|l|l|l|l|l||}\hline
Observables & Measured value &Fit &Model
&Fit &Model  \\ \hline
&&\multicolumn{2}{l|}{$M_H=100$ GeV} & \multicolumn{2}{l||}
{$M_H=400$ GeV}
\\ \hline
$\Gamma_Z(\hbox{GeV})$ & $2.4963\pm 0.0032$ & $2.500$ & $2.501$ & 
$2.499$ & $2.500$
\\ \hline
$R_{\ell}$ & $20.788\pm 0.032$ & $20.76$ & $20.78$ & $20.76$ & $20.78$  
\\ \hline
$\sigma_h^0(\hbox{nb})$ & $41.488\pm 0.078$ & $41.46$ & $41.45$ 
& $41.45$ & $41.44$
\\ \hline
$R_b$ & $0.2219\pm 0.0017$ & $0.2200$ & $0.2195$ & $0.2210$ & $0.2205$  
\\ \hline
$R_c$ & $0.1540\pm 0.0074$ & $0.1642$ & $0.1649$ & $0.1626$ & $0.1634$  
\\ \hline
$A^b_{FB}$ & $0.0997\pm 0.0031$ & $0.1043$ &$ 0.1044$ & $0.1017$ &$ 0.1013$  
\\ \hline
$A_b(LR)$ & $0.841\pm 0.053$ & $0.9284$ &$ 0.9297$ &$ 0.9285$ &$ 0.9281$  
\\ \hline
$A^c_{FB}$ & $0.0729\pm 0.0058$ &$ 0.0784$ &$ 0.0775$ &$ 0.0766$ &$ 0.0757$  
\\ \hline
$A_c(LR)$ & $0.606\pm 0.090$ & $0.698$ & $0.690$ & $ 0.699$ &$ 0.693$  
\\ \hline
$A_{\tau}$ & $0.1418\pm 0.0075$ & $0.1497$ & $0.1497$ &$ 0.1461$ 
& $0.1456$  \\ \hline
$A_e$ &$ 0.1390\pm 0.0089$ & $0.1497$ & $0.1497$ & $0.1461$ & $0.1456$  
\\ \hline
$A_{LR}$ &$ 0.1551\pm 0.0040$ &$ 0.1497$ &$ 0.1497$ &$ 0.1461$ &$ 0.1456$  
\\ \hline
$A^{\ell}_{FB}$ &$ 0.0172\pm 0.0012$ &$ 0.0169$ &$ 0.0169$ &$ 0.0161$ 
&$ 0.0159$  
\\ \hline
$Q_W(Cs)$ &$ -71.04\pm 1.81$ &$ -70.74$ &$ -71.78$ &$ -71.03$ &
$-71.36$  \\ \hline
$Q_W(Tl)$ &$ -114.2\pm 3.8$ &$ -113.0$ &$ -114.7$ &$ -113.5$
& $-114.0$  \\ \hline
$g_L^2$ & $0.2980\pm 0.0044 $ &$ 0.300$ &$ 0.300$ &$ 0.3007$ &$ 0.3004$
\\ \hline
$g_R^2$ & $0.0307\pm 0.0047 $ &$ 0.0279$ &$ 0.0274$ &$ 0.0269$ &$ 0.0274$
\\ \hline
$M_W$ (GeV) & $80.33\pm 0.15 $ &$ 80.43$ &$ 80.43$ &$ 80.38$ &$ 80.37$
\\ \hline
$\chi^2$ & $$ & $16$ & $17$ & $15$ & $16$  \\ \hline
\end{tabular}
\end{center}
\label{tble1}
\caption{Experimental 
\protect\cite{lepdata,slac,cesium,thalium,wmass,neutrino}
and fitted values of observables. Correlations between the data were
included in the fit. Column labeled ``Fit'' shows fitted values of 
observables for arbitrary $U(1)_X$
quark charges (14 d.o.f.). Column labeled ``Model'' gives fitted 
values for the model discussed in
the text (16 d.o.f.). All fitted values are for $g_X=0.15$. 
The $\chi^2$ for the Standard Model is
30 (18 d.o.f.) for $M_H=$ 100 GeV.}
\end{table}

A model looks more natural if the $U(1)_X$ charge
ratios are rational numbers. Though this is by no means a requirement, 
we restrict to this possibility.  
We find that the fitted charges can accommodate such a choice:  
$q_Q = -1$, $q_U = 2$ and $q_D = 1$. Three Higgs doublets $H_u$, $H_d$ 
and $H_l$ are introduced to give masses to the up 
quarks, down quarks and leptons.  Their  $U(1)_X$ charges are 
$q_{H_u}=-3$, $q_{H_d}=-2$ and $q_{H_l}=0$.
Since the $U(1)_X$ charges of the Standard Model fermions are 
not vector-like, 
the $U(1)_X$ gauge symmetry is anomalous and new quarks must be added 
to cancel the anomalies. We add
three generations of Standard Model-like quarks with opposite 
$U(1)_Y$ and $U(1)_X$ charges: 
$ Q'_L$ = ({\bf 2}, -1/6, 1), $\tilde{U}_R$ = ({\bf 1}, -2/3, -2), 
$\tilde{D}_R$ = ({\bf 1}, 1/3, -1) under $SU(2)\times U(1)_Y\times U(1)_X$.  
These new quarks, in turn, generate anomalies under 
$SU(2)\times U(1)_Y$ and
their chiral partners must be added to make 
$SU(2)\times U(1)_Y$ vector-like: 
$ Q'_R$ = ({\bf 2}, -1/6, 0),
$\tilde{U}_L$= ({\bf 1}, -2/3, 0), $\tilde{D}_L$= ({\bf 1}, 1/3, 0). 
Since all of the added quarks are vector-like under $SU(2)\times U(1)_Y$, 
there is no contribution to the S parameter \cite{s-param}. Since twelve 
quark flavors have been added to cancel anomalies, the QCD coupling 
is no longer 
asymptotically free. Using the one-loop $\beta$
function, we have checked that the coupling remains perturbative 
up to the Planck scale.

The new quarks should be heavier than about 200 GeV to avoid detection at 
Fermilab. They can acquire mass through the Higgs doublets:
\begin{equation}
       \lambda_1 \overline{Q'}_L\tilde{U}_R H_u^c  
     + \lambda_2  \overline{Q'}_R\tilde{D}_L H_l^c 
     + \lambda_3 \overline{Q'}_L\tilde{D}_R H_d^c
     + \lambda_4 \overline{Q'}_R\tilde{U}_L H_l,   \label{doublet}
\end{equation} 
where the superscript $c$ denotes charge conjugation. The masses 
generated by these couplings should be of the order of the Standard 
Model quark masses.
To make the new quarks heavier, we must introduce 
additional singlet Higgs couplings. These singlet Higgs fields break
$U(1)_X$ at a scale larger than the electroweak scale.  
Two Higgs singlets
$\phi$ and $\phi^\prime$ are introduced with the $U(1)_X$ 
charges $q_{\phi}=-1$ 
and $q_{\phi'}=-2$ so that the new quarks acquire mass through
\begin{equation}
    \lambda_5 \overline{Q'} _R  Q'_L \phi +   
     \lambda_6 \overline{\tilde{U}}_R  \tilde{U}_L \phi^{\prime} 
    +\lambda_7 \overline{\tilde{D}}_R  \tilde{D}_L \phi  
\label{singlet}    
\end{equation} 
When contribution of Eq.(\ref{singlet}) is much larger than that of 
Eq.(\ref{doublet}), the new quarks are nearly degenerate and a 
shift in the 
T parameter \cite{s-param} is negligible.

Since five Higgs fields ($H_u, H_d, H_l, \phi$ and $\phi^\prime$) 
develop  
{\it vevs}, we must ensure that they do not result 
in an unabsorbed Goldstone boson 
or an axion. We introduce self-interactions among the Higgs 
multiplets to
eliminate accidental global symmetries that may break 
down spontaneously.
Since two neutral gauge bosons of 
$SU(2)\times U(1)_Y\times U(1)_X$ absorb 
two Goldstone modes, we add appropriate interactions among 
Higgs fields to
give mass to the three remaining modes. The following couplings suffice:
\begin{equation}
     \lambda_8 \phi^2 \phi^{\prime c}  +  
     \lambda_9 H_u^c H_l^c \phi \phi^{\prime}  +
     \lambda_{10}  H_d H_l^c \phi^{\prime c}                
\end{equation} 

%Another possible phenomenological problem is mixing between the Standard
%Model quarks and the new quarks. If $(Q'_L, \tilde{U}_R, \tilde{D}_R)$ 
%were
%assigned to color triplets, the lightest new quark would be stable and
%the lightest baryonic bound state of the new quarks might be abundant 
%enough
%to have been detected in exotic matter searches \cite{smith}.  

The new quarks can be 3 or $\bar{3}$ of $SU(3)$. 
If $(Q', \tilde{U}, \tilde{D})$ are
assigned to color triplets, there is an accidental discrete symmetry, 
$Q \to  Q $, 
$Q' \to  - Q'$ etc., that prevents the lightest new quark from decaying. 
Then
the lightest baryonic bound state of the new quarks might be abundant 
enough
to have been detected in exotic matter searches \cite{smith}.
When they are assigned to color antitriplets, we can introduce another 
scalar singlet
$\tilde{\phi}$ and allow the new quarks to decay 
into $Q$ and $\tilde{\phi}$ through the coupling $QQ'\tilde{\phi}$.
However, the following mass terms are then allowed by the gauge 
symmetries:  
\begin{equation}
       M_1 Q_L^T C Q'_L +  M_2 U_R^T C \tilde{U}_R + 
M_3 D_R^T C \tilde{D}_R,
\end{equation}
where $C$ is the charge conjugation matrix.
These terms result in mixing between the Standard Model quarks and
the new quarks.
They may be forbidden 
by imposing the discrete symmetry mentioned above. 
We assign an odd parity to $\tilde{\phi}$ under this symmetry to
maintain the $QQ'\tilde{\phi}$ coupling.
Since $\tilde{\phi}$ is a singlet 
carrying no $U(1)_Y$ or $U(1)_X$ charge and is stable, it can escape 
detection in 
terrestrial experiments.  
The $\tilde{\phi}$ particle could have been produced in the early Universe 
and could contribute to the mass density.  The mass and coupling of 
$\tilde{\phi}$ can be adjusted so that it does not overclose the 
Universe \cite{cdm}. 

\begin{table}
\begin{center}
\begin{tabular}{||l|l|l|l|l|l||}\hline
Fields &$SU(3)$&$SU(2)$&$U(1)_Y$&$U(1)_X$\\ \hline
$Q_L$&{\bf 3}&{\bf 2}&1/6&-1 \\ \hline
$U_R$&{\bf 3}&{\bf 1}&2/3&+2 \\ \hline
$D_R$&{\bf 3}&{\bf 1}&-1/3&+1 \\ \hline
$Q'_L$&${\bf \bar{3}}$&{\bf 2}&-1/6&+1 \\ \hline
$\tilde{U} _R$&${\bf \bar{3}}$&{\bf 1}&-2/3&-2 \\ \hline
$\tilde{D} _R$&${\bf \bar{3}}$&{\bf 1}&+1/3&-1 \\ \hline
$Q'_R$&${\bf \bar{3}}$&{\bf 2}&-1/6&0 \\ \hline
$\tilde{U} _L$&${\bf \bar{3}}$&{\bf 1}&-2/3&0 \\ \hline
$\tilde{D} _L$&${\bf \bar{3}}$&{\bf 1}&+1/3&0 \\ \hline
$H_u$&{\bf 1}&{\bf 2}&-1/2&-3 \\ \hline
$H_d$&{\bf 1}&{\bf 2}&+1/2&-2 \\ \hline
$H_l$&{\bf 1}&{\bf 2}&+1/2&0 \\ \hline
$\phi$&{\bf 1}&{\bf 1}&0&-1 \\ \hline
$\phi ^{\prime}$&{\bf 1}&{\bf 1}&0&-2 \\ \hline
$\tilde{\phi}$&{\bf 1}&{\bf 1}&0&0 \\ \hline
\end{tabular}
\end{center}
\caption{$SU(3) \times SU(2) \times U(1)_Y \times U(1)_X$
quantum numbers for matter fields in our model. }
\end{table}

We now examine the property of the $X$ boson in our model and some of its
phenomenological implications. The parameter $\kappa$ in the 
$Z-X$ mass matrix
is given by
\begin{equation}
        \kappa = \frac{2v_d^2 - 3v_u^2}{2v^2},
\end{equation}
where $\vev{H_u} = v_u/\sqrt{2}$, $\vev{H_d} =v_d/\sqrt{2}$ and
 $\vev{H_l} = v_l/\sqrt{2}$ with \\
$\sqrt{v_u^2 + v_d^2 + v_l^2} = v = 247$
GeV. Introducing $\vev{\phi}=V/\sqrt{2}$, $\vev{\phi'}= V'/\sqrt{2}$ and
$5\tilde{V}^2 \equiv V^2+4V^{{\prime}{2}}$, the parameter $V_X^2$ 
is given by $5\tilde{V}^2 + 9v_u^2 + 4v_d^2$. Assuming
$v_l^2 \ll v^2$, the parameters to be fitted are $g_X$, $\kappa$ and 
$\tilde{V}$.
If the Yukawa couplings appearing in Eq.(\ref{singlet}) are O(1), 
the {\it vevs}
 $V$ and $V'$ should be greater than $v$ so that the new quarks 
are heavier than
the Standard Model quarks. This implies a lower limit $\tilde{V} 
\ge 250$ GeV
(or equivalently $V_X$ $\gtap$ $750$ GeV) that is imposed in 
performing the fit. 
We restrict $g_X \geq 0.1$ to guarantee $M_X$ $>$ $M_Z$. Since the 
charges of the up quarks under 
$U(1)_X$ are large, we restrict $g_X \leq 0.5$ to ensure that 
the coupling strength 
of the $X$ boson to up quarks $(g_L^2+g_R^2)g_X^2/4\pi=5g_X^2/4\pi$ 
remains perturbative.
                
We now discuss a fit to the 18 observables using $M_H=400$ GeV. 
Diagonalizing the mass matrix
exactly (which gives the mass of the X boson) in this model gives 
$\chi^2=16$, 
$\tilde{V} = 340$ GeV, and $\kappa =-0.05$ 
for $g_X = 0.15$ fixed. The $\chi^2$ is not very
sensitive to $g_X$. The $\chi^2$ increases by $1.5$ if $g_X$ is 
varied between 
$0.1$ and $0.5$. With
$g_X=0.15$ fixed, the 90\% (95\%) C.L. range for $\tilde{V}$ is 
250 GeV to 980 GeV (250 GeV to 1200 GeV) and the 90\% (95\%) 
C.L. range for $\kappa$ is  -0.01 to -0.18 (-0.01 to -0.23). 
The allowed range for $\kappa$ is -1.5 to 1. Therefore $\kappa$ 
must be fine
tuned to within 6.8 (8.8) \%. 

The mass of the X boson depends on $\kappa$, $\tilde{V}$ and
(almost linearly) on $g_X$. For small $\kappa$, $M_X^2\simeq 
g_X^2(5\tilde{V}^2+6v^2)$. The 
90\% (95\%) C.L. range for $M_X$ is from 125 GeV to 341 GeV 
(125 GeV to 412 GeV) for $g_X=$0.15 and 
$M_H=400$ GeV.
For $g_X=0.5$ and $M_H=100$ GeV, the 90\% (95\%) C.L. range for 
$M_X$ is from 420 GeV 
to 820 GeV (420 GeV to 1070 GeV).
The $X$ boson can be produced in $p\overline{p}$ collisions at the 
Tevatron and
detected in the dijet final state. For $g_X=0.15$, the expected 
production rate
is considerably below the limit set by the CDF group \cite{cdfdijet} 
for all
values of $\tilde{V}$. For $g_X=0.5$, the values of 
$M_X \ltap$  750 GeV \\
($\tilde{V} \ltap
$ 610 GeV) are excluded. For $g_X=0.3$, the region $320$ GeV 
$ \ltap M_X \ltap 
$ 520 GeV is excluded.

    To summarize, on the basis of an overall fit to all electroweak data 
we have built a viable $U(1)$ extension of the Standard Model. While 
the fit to
data has been greatly improved, the model lacks aesthetic appeal.
The $X$ boson may be accessible by the experiments at 
Fermilab in the future.

K.A. and M.G. would like to thank Nima Arkani-Hamed and Chris Carone 
for useful
comments. The work was supported in part by the Director, Office 
of Energy Research, 
Office of High Energy Physics, Division of High Energy Physics of the 
U.S. Department of Energy under Contract DE--AC03--76SF00098 and in part 
by the National 
Science Foundation under grant PHY-90-21139.
Accordingly, the U.S. Government retains a nonexclusive, royalty-free 
license to publish or reproduce the published form of this contribution, 
or allow others to do so, for U.S. Government purposes.

\end{document}